\documentclass[sigconf]{acmart}

\AtBeginDocument{%
  }

\copyrightyear{2026}
\acmYear{2026}
\setcopyright{cc}
\setcctype{by}
\acmConference[C\&C '26]{Creativity and Cognition}{July 13--16, 2026}{London, United Kingdom}
\acmBooktitle{Creativity and Cognition (C\&C '26), July 13--16, 2026, London, United Kingdom}
\acmDOI{10.1145/3803784.3807570}
\acmISBN{979-8-4007-2583-8/2026/07}

\usepackage{subcaption}
\usepackage{soul}

\begin{document}

\title[How Creatives Approach GenAI Image Generation]{How Creatives Approach GenAI Image Generation: Tensions Between Structured Guidance, Self-Experimentation, and Creative Autonomy} 

\author{Haidan Liu}
\affiliation{%
  \department{Computing Science}
  \institution{Simon Fraser University}
  \city{Burnaby}
  \country{Canada}}
\email{haidanl@sfu.ca}

\author{Isabelle Kwan}
\affiliation{%
  \department{Computing Science}
  \institution{Simon Fraser University}
  \city{Burnaby}
  \country{Canada}}
\email{itk1@sfu.ca}

\author{Taiga Okuma}
\affiliation{%
  \department{Computing Science}
  \institution{Simon Fraser University}
  \city{Burnaby}
  \country{Canada}}
\email{toa12@sfu.ca}

\author{Jeffrey Loverock}
\affiliation{%
  \department{Computing Science}
  \institution{Simon Fraser University}
  \city{Burnaby}
  \country{Canada}}
\email{jla956@sfu.ca}

\author{Nicholas Vincent}
\affiliation{%
  \department{Computing Science}
  \institution{Simon Fraser University}
  \city{Burnaby}
  \country{Canada}}
\email{nvincent@sfu.ca}

\author{Parmit K Chilana}
\affiliation{%
  \department{Computing Science}
  \institution{Simon Fraser University}
  \city{Burnaby}
  \country{Canada}}
\email{pchilana@sfu.ca}

\renewcommand{\shortauthors}{Liu et al.}

\begin{abstract}

As generative AI tools increasingly influence creative practice, they raise longstanding HCI questions about how creatives learn complex software and how they can be better supported. We conducted an interview study with artists and hobbyists (n=8) and a follow-up survey (n=159) to understand how this population approaches and seeks guidance for GenAI image tools. We found that creatives commonly use either self-experimentation or tutorials to explore GenAI tools, yet many struggle with confusing AI terminology. To gain further insight into creatives' learning experiences, we developed a research probe to elicit creatives' perceptions of structured guidance. Our user study with 17 creatives revealed that, even when creatives described the guidance as helpful for understanding AI, many still preferred self-experimentation, feeling that guidance could limit their creativity. Our findings highlight a central tension in supporting AI literacy for creatives: balancing guidance and promoting literacy while preserving creative freedom. 

\end{abstract}

\ccsdesc[500]{Human-centered computing~Human computer interaction (HCI)}

\keywords{Generative AI Image Tools, Mental Models, AI Literacy}
\maketitle

\section{Introduction}

Generative Artificial Intelligence (GenAI) image tools (e.g., \textit{DALL·E} \cite{dalle}) and mobile applications (e.g., \textit{WhatsApp} \cite{whatsup}, \textit{Instagram} \cite{ins}) have lowered barriers to AI-assisted creativity by enabling image generation from natural language prompts \cite{dilodovico}. However, their ``black box'' nature leaves users with limited understanding of how prompts are interpreted, making it difficult to refine outputs, build reliable workflows, or form useful mental models \cite{black,dilodovico,15}. This opacity also creates challenges for organizations trying to improve AI literacy, or makes it harder for users to become AI literate by limiting users' ability to evaluate outputs and engage critically with AI \cite{static, Burgsteiner, ghallab, dilodovico}.

Recent GenAI tools increasingly support output steering through interface-level controls, such as localized editing, richer prompting workflows, and production-oriented features \cite{editing, adobeimage}. While these features improve usability and reduce friction, they primarily support outcome control rather than conceptual understanding. Users may gain more ways to manipulate results, but still lack transferable intuitions about how systems interpret inputs, why outputs vary, or when failures occur. As models change over time, users must also continually adapt strategies that may not transfer across versions or tools \cite{gagan, aiwildcard, oppenlaender2024prompting}.

\begin{figure*}[t]
    \centering
    \includegraphics[width=0.9\linewidth]{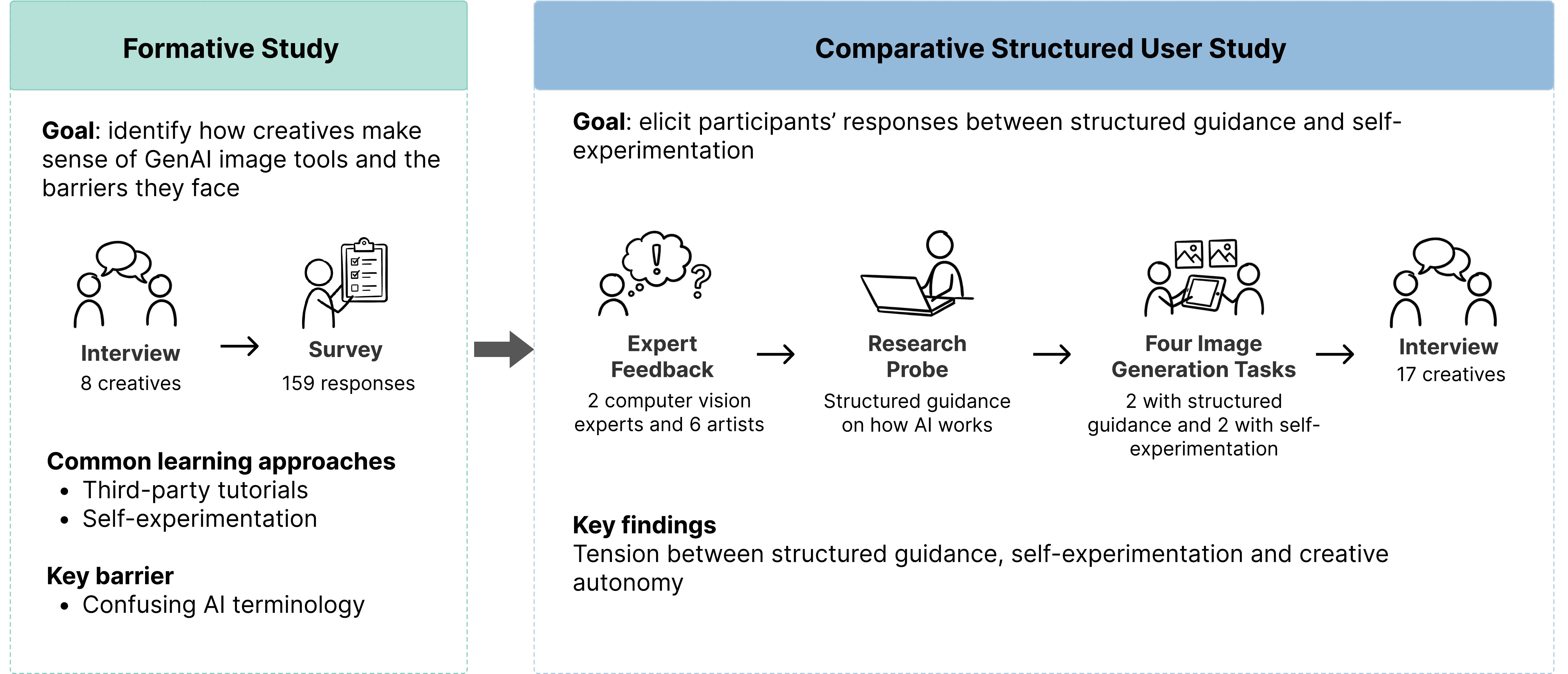}
    \caption{Overview of our staged inquiry across three studies. The formative interviews and survey studies (left) identified creatives' common learning strategies and challenges in making sense of GenAI image tools. We then developed and refined a research probe that provided easy-to-understand guidance on how GenAI works. The comparative structured user study (right) using the research probe to elicit participants' perceptions on structured guidance and self-experimentation and surfaced the tension between structured support and creative autonomy. The overall research process spanned summer 2024 to summer 2025.}
    \label{fig:methodprocess}
\end{figure*}

We adopt a mental models lens \cite{norman1983} to examine how creatives understand and reason about GenAI image tools. 
Throughout this paper, we use the term ``creatives'' to refer to visual artists and hobbyists who engage with GenAI image tools in creative contexts. We first conducted interviews with 8 visual artists and hobbyists to understand how they approach GenAI image tools and the challenges they face when trying to make sense of how AI interprets their inputs and generates outputs. While most preferred self-experimentation, some turned to tutorials and expressed a desire for simpler, visual explanations. 
To assess how prevalent these patterns are beyond our interview sample, we followed up with an online survey with 159 visual artists and hobbyists to assess their views on existing tutorials and support resources for popular GenAI image tools. Our survey findings revealed that third-party online tutorials (e.g., from YouTubers and other content creatives or online courses rather than those from tool platforms) are the most popular strategies among creatives when approaching GenAI image tools. Respondents showed greater interest in tutorials that help them understand AI concepts and parameter settings though complex terminology emerged as a key barrier in the process (Figure~\ref{fig:surveyfindings}b).

Building on these formative findings, we next sought to understand how creatives respond to simplified structured guidance when technical barriers such as jargon and overly complex guidance are minimized. We use ``structured guidance'' to refer to learning support, such as tutorials, videos, walkthroughs, examples, or in-tool explanations, that organizes information to help users understand and use GenAI tools. Existing guidance formats were not well-suited for this purpose as many are highly technical and rely on specialized terminology that can be difficult for non-expert users to follow (e.g., \cite{nvida, jay}), while others are tailored to specific applications (e.g., \cite{nightcafework}). To address this gap, we created our own \textit{research probe} \cite{designprobe} to elicit participants' responses on structured guidance of how AI works without the confounds introduced by these existing guidance formats (e.g., technical jargon). We then conducted a comparative structured observation study and follow-up interviews with 17 visual artists and hobbyists to elicit reflections on their experience with the probe during image-generation tasks and how they balanced structured guidance with self-experimentation. 

Our findings show that structured guidance can help users by clarifying how the AI interprets prompts and decides what to try next for better results. We also observed an intriguing paradox: while structured guidance supported understanding, some creatives still preferred self-experimentation, describing it as a better match for preserving creative autonomy during exploration. These dynamics suggest that supporting creative work with GenAI tools requires a closer understanding of how creatives build mental models through interaction, while also respecting their desire to maintain creative autonomy as they navigate both AI behavior and external guidance. Our study offers insights into these processes and highlights the need for learning resource designs that foster conceptual understanding of AI systems without diminishing creative freedom.

Across our studies, we contribute a staged account of how creatives learn GenAI tools. Our formative interviews identified two dominant learning approaches of self-experimentation and tutorials and a key barrier: difficulty with technical terminology, alongside a demand for simpler, conceptual explanations. The survey showed that these patterns persist at scale, confirming both the prevalence of these strategies and dissatisfaction with existing tutorials. Building on this, the probe study isolates simplified structured visual guidance and examines how creatives engage with it in practice relative to self-experimentation, showing that while guidance can improve understanding of how the system interprets inputs, many creatives still prefer self-experimentation to maintain creative autonomy, that is choosing one's creative path and expressing unique ideas \cite{klein1989integrated,rafner2025agency}. We argue that ``more tutorials'' is not the solution. Learning support should be situated, optional, and adaptable, offering in-context explanations that respect creative agency, the creator’s ability to shape the creative process and outcomes \cite{beghetto1945creative,rafner2025agency}, and reflect that creators’ literacy needs are uneven, personal, and goal-driven.

\section{Related Work}

We draw on four strands of prior work: AI literacy and creative education, learnability of GenAI tools, co-creation with GenAI tools, and mental models for understanding creatives' interactions with GenAI tools.

Before reviewing the literature, we clarify several terms used throughout the paper. We use machine learning (ML) to refer to systems that detect patterns in data \cite{tseng2023}; non-ML-expert users are users without machine learning expertise. We use GenAI to refer to systems that generate new content from user input \cite{muller2023}, with this paper focusing on GenAI image tools. Large language models (LLMs) are text-based GenAI systems trained on large-scale language data \cite{naveed2025}.

\subsection{AI Literacy in Creative Context}
\citet{durilong20} describe AI literacy as ``a set of competencies that enables individuals to critically evaluate AI technologies; communicate and collaborate effectively with AI; and use AI as a tool online, at home, and in the workplace.'' Researchers have proposed a range of AI literacy frameworks to guide educational and design interventions, mostly in computing education contexts, often overlapping in understanding AI \cite{ng, ngb,druga2019}, applying AI \cite{druga2019, druga2022}, evaluating AI \cite{durilong20, wong2020,ngb}, and AI ethics \cite{wong2020, ng}. Conceptual understanding and foundational knowledge are often treated as core prerequisites for being considered ``AI literate'' \cite{ng, druga2019, druga2022}. Recent work on online creative communities suggests that creatives often prioritize practical AI literacy, focusing on how to use tools effectively \cite{liu2026tracing}.

Work in creative education reflects many of the same priorities emphasized in emerging AI literacy frameworks. Researchers in media and design education argue that creatives should understand how AI works and recognize its ethical implications, rather than adopting AI tools in a passive way \cite{wolf2025}. Research in design pedagogy similarly highlights the importance of integrating AI competencies into the curriculum to prepare students for AI-supported creative industries \cite{fleischmann, fleischmann2024}. These competencies span technical understanding of AI, critical appraisal of AI-generated outputs, and practical application skills \cite{schauer2024}.  

Although conceptual understanding is often treated as central to AI literacy, recent work on online creative communities suggests that creatives often prioritize practical AI literacy, focusing on how to use tools effectively.

\subsection{Learning to Use GenAI Tools}

Learning to navigate complex, feature-rich software is a long-standing challenge in HCI \cite{gi21}. Experiential learning theory argues that skills develop through iterative cycles of action, reflection, and revision rather than only through upfront instruction \cite{kolb2014}. In GenAI image tools, this dynamic often plays out through prompting, which has emerged as a creative skill shaped by systematic trial and error and repeated refinement \cite{promptartist}. For example, creatives learn through trial and error with different models, prompts, and settings to understand the tool's capabilities and limitations in AI-based manufacturing design \cite{51, wild}. Prior work also shows that creatives build practices and community knowledge by exploring, reusing, and iteratively adapting prompts over time \cite{promptartist}. 

However, because GenAI tools provide limited transparency into how training data, randomness, and interface parameters jointly shape outputs \cite{oppenlaender2022, oppenlaender2024prompting}, trial and error can also foster over-simplified or inaccurate mental models of system behavior \cite{anjaliiui}. In music generation, prior work has supported non-ML-expert users by mapping latent dimensions to musical attributes and providing real-time feedback and visualizations \cite{banar}. Recent work in image generation embeds explainability into node-based diffusion workflows, supporting tacit understanding through component-level manipulation \cite{abuzuraiq2025explainabilityinaction}. 

At the same time, guidance introduces its own design tension. Prior work on computational feedback in visual design tools shows that actionable scaffolding can help novices apply design principles and explore alternatives, but may also narrow exploration and increase the risk of overreliance \cite{viz2026}. Building on this interest, we explore how creatives without ML expertise with prompt-based commercial GenAI image tools perceive simple conceptual explanations.

\begin{figure*}[t]
    \centering
    \includegraphics[width=0.7\linewidth]{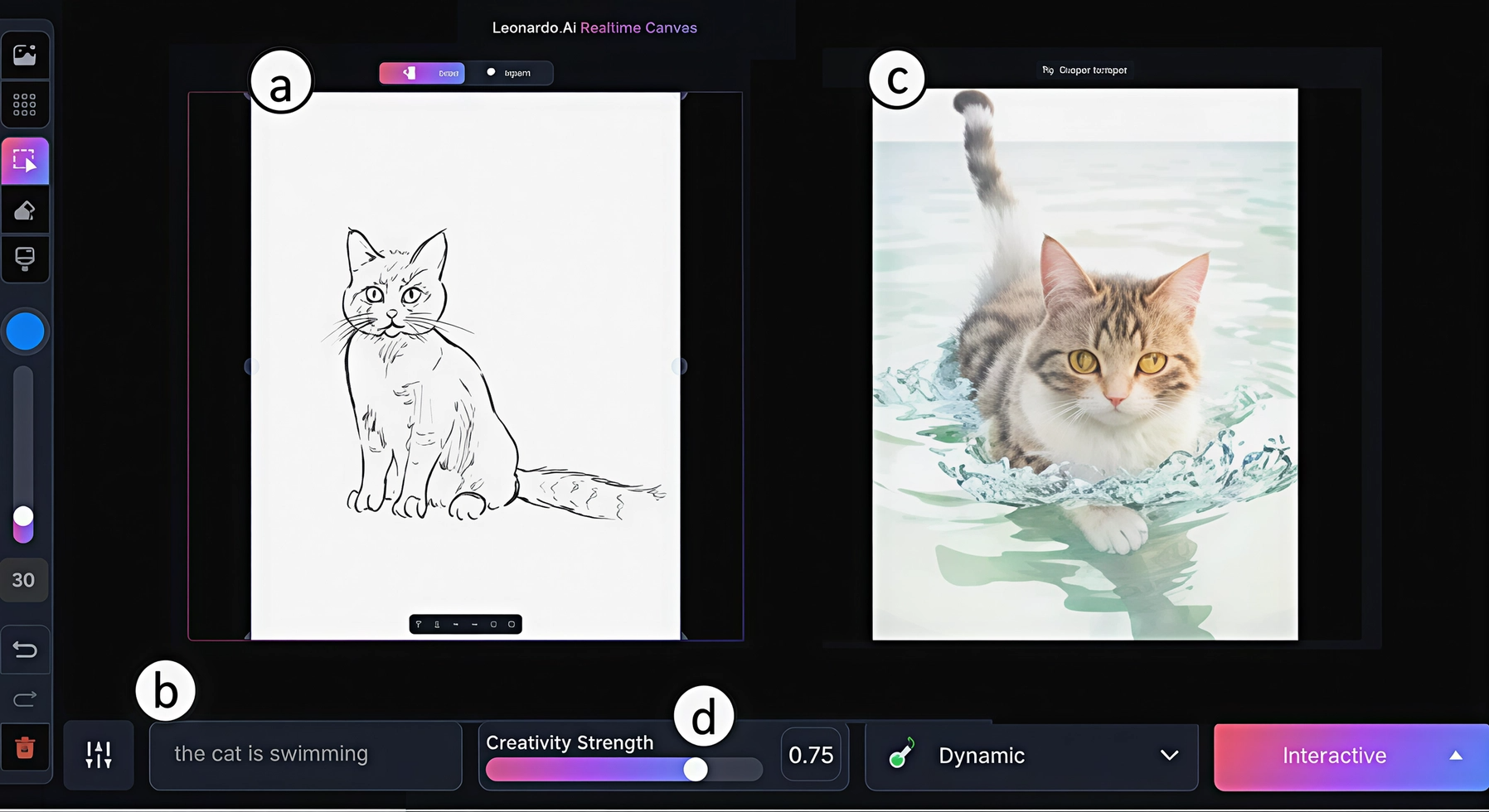}
    \caption{Leonardo AI interface used in the interview study. Participants were given a sketch image provided by the researchers (a), then entered a short text prompt describing the intended image (b), and viewed the generated result (c). They could further adjust the outcome using the creativity strength parameter (d), which controlled how much the result diverged from the original input.}
    \vspace{-6pt}
    \label{fig: leonardointerface}
\end{figure*}

\subsection{Co-creation with GenAI tools}

Recent studies show that creatives increasingly use GenAI image tools for ideation, automation, and visual refinement, but the impact of these tools varies across creative roles and domains \cite{late, 16, disdesignprompt}. While hobbyists may embrace GenAI for expanding creative possibilities, professionals, such as illustrators and animators, face growing concerns around deskilling, task automation, and shifting creative authority \cite{chang2024impact}. Yet, day-to-day use remains challenging: prompt crafting can be slow and frustrating \cite{inefficient, 34, 15}, users may struggle with articulating their creative goals without specialized vocabulary \cite{evolving}, and users are often unsure whether misaligned results reflect limitations in their input or in the model \cite{journey, disInfuser}. 

In response, HCI research has proposed ways to support prompt creation and refinement \cite{pixel, prompt_this, taxonomy, promptify}. Some contributions offer reusable vocabulary and strategies, such as taxonomies of prompt modifiers \cite{taxonomy}; others provide interactive scaffolding through LLM-based suggestions \cite{promptify} or visualized prompt-editing traces \cite{prompt_this}. Interfaces like PromptPaint enable regional, direct-manipulation prompting \cite{chung_promptpaint_2023}. Existing support tools are primarily early-stage prototypes that focus on improving iteration and control \cite{kim2024}, but they offer less support for helping creatives understand misaligned outputs or how guidance shapes their next-step strategies. In this paper, we contribute by studying how creatives interpret AI behavior, understand prompt-output relationships, and balance structured guidance with creative autonomy.

\subsection{Mental Models as a Lens for Understanding GenAI Tools}

Norman~\cite{norman1983} argues that people develop functional mental models through interaction with a system to guide their actions and interpret outcomes. Research shows that users of complex and opaque systems often develop partial or inaccurate models that shape their problem-solving strategies, errors, and sense of control \cite{kulesza2012}. Without clear cues about how a system works, users frequently fill these gaps with simplified ``folk'' theories of system behavior \cite{dilodovico}, which can support basic interaction but produce brittle expectations and ineffective strategies \cite{kulesza2012,bansal2019,kocielnik2019}. 

These difficulties align with \textit{Gulf of Execution} and \textit{Gulf of Evaluation}: the gaps between what users intend to do and what the system makes possible, and between what the system does and how easily users can interpret it \cite{norman1983}. Recent work has extended these concepts to AI systems, identifying alignment challenges around specification, process, and evaluation \cite{terry2023}. Relatedly, studies of LLM interactions, where users communicate with text-based GenAI systems through natural language prompts and interpret generated text responses, characterizing a ``Gulf of Envisioning'' in which users struggle to understand system capabilities, articulate intentions, and anticipate outputs \cite{bridge2024}. Providing structured information may help bridge these gulfs by supporting clearer action-outcome mappings and helping users align their mental models with the system's actual behavior \cite{norman1983}.

Our work uses the lens of mental models to examine how creatives make sense of genAI image tools. We also compare structured guidance with self-experimentation to reveal how each pathway cultivates distinct mental models of how the system works.

\section{Formative Study: How Artists and Hobbyists Approach GenAI Image Tools}

To understand how creatives form mental models of GenAI image tools and the kinds of support they find meaningful, we first conducted an interview study with 8 artists and hobbyists. We then distributed an online survey across art-focused communities to examine how these patterns appear at scale, yielding 159 responses. 

\subsection{Interview Study}

The primary goal of our interview study was to examine how creatives approach learning an unfamiliar GenAI image tool, including their strategies and challenges.

\textbf{Participants and Procedure:} We recruited eight participants with diverse creative backgrounds and demographics. Three participants identified as artists, while five described themselves as hobbyists. Five sessions were conducted in person, and three sessions were held online through Zoom. Each session lasted approximately 45 minutes, and participants received a \$20 Amazon gift card as compensation for their time.

\begin{figure*}[t]
    \centering
    \includegraphics[width=0.92\linewidth]{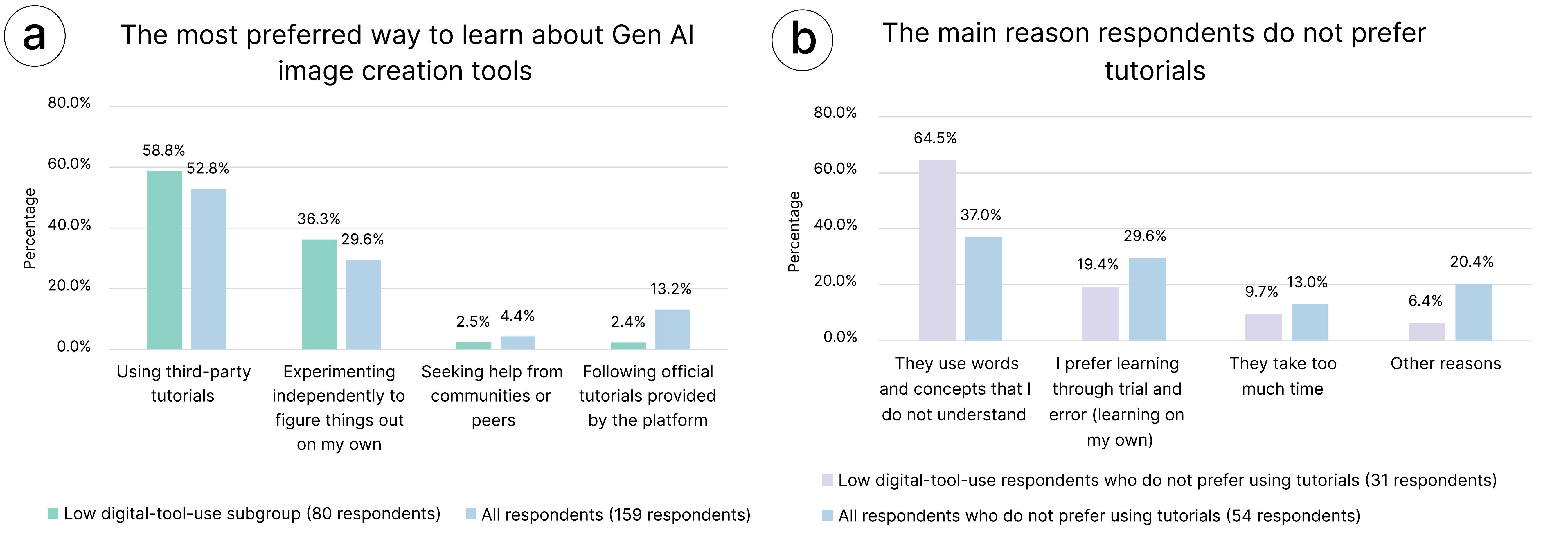}
    \caption{(a) Responses to the survey question (single-select): What is your most preferred way to learn about GenAI image creation tools (e.g., DALL-E, Midjouney, Stable Diffusion, Artbreeder, Runway ML, or others)? The top learning strategy for GenAI image tools was following third-party tutorials, with self-experimentation close behind. The low digital-tool-use subgroup was defined based on answers to the question ``How often do you use digital tools or software (e.g., Procreate, Photoshop, Illustrator) in your art practice?'' Respondents who selected Never, Rarely, or Sometimes were included in this subgroup (n = 80). (b) Responses to the question (single-select): If you don't prefer using tutorials, what is the main reason? Tutorials were often avoided due to confusing terminology, a preference for self-experimentation, and the perception that they're time-consuming. Other reasons include difficulty following them, limited depth, or simply a lack of interest—like one user who preferred making art solo.}
    \label{fig:surveyfindings}
\end{figure*}

During the session, participants explored the \textit{Leonardo AI} \cite{leonardoai} image generation tool, selected for its easy setup and range of features. We gave participants three suggested tasks to elicit their reactions and feedback: generating images from text prompts (text-to-image), training a model on example images to create new ones (image-to-image), and transforming hand-drawn sketches into detailed images (sketch-to-image). Before each task, a researcher briefly introduced the interface. Participants chose which tasks to try, explored at their own pace, and then completed brief semi-structured interviews about their approaches and challenges.

We analyzed interview transcripts inductively \cite{strauss1994grounded}, using open coding \cite{benaquisto_open_2008} to iteratively develop five analytic categories around participants' learning experiences and challenges. We then used thematic analysis \cite{thematic} to synthesize these categories into the two overarching themes reported in the Key Findings.

\textbf{Key Findings:} Our interview study surfaced two recurring themes in how participants engaged with GenAI image tools: reliance on self-directed trial and error and persistent terminology confusion. Across participants, learning was largely driven by self-experimentation, which helped them form initial—though often partial—understandings of how the tools behaved. As P08 explained, \textit{``I just want to click around \ldots I feel like AI is pretty straightforward.''} Similarly, P05, who used GenAI to design posters for his coffee shop, described his approach as \textit{``all try and fail,''} noting that despite the time and cost involved, experimenting on his own was still preferable to hiring external help.

Yet, most participants expressed a desire for deeper conceptual understanding of how GenAI systems interpret prompts, revealing gaps in their mental models. Several had attempted tutorials but found them unsatisfying. P06, with a background in geographic information science, said she wanted to \textit{``vaguely know how the model works''} through visual explanations, such as illustrations or videos, but struggled to find accessible resources. P07, who did not have a technical background, similarly sought to understand AI through YouTube tutorials but found many overly technical, explaining that she wanted \textit{``very simple explanations''} that were visual rather than text-heavy.

Participants also reported difficulty interpreting specialized terminology in GenAI interfaces, which often left them uncertain about what the system was doing or how to proceed. Terms such as \textit{``training dataset''} or \textit{``model description''} were frequently described as confusing. As P04 noted, \textit{``I don't know what half of these things mean, so I just type in the box that looks like it's for typing.''} Others highlighted the need for clearer explanations of parameters: P08 remarked that understanding what the \textit{``creativity strength''} setting represented \textit{``would be very helpful.''}

\begin{figure*}[t]
    \centering
    \includegraphics[width=0.92 \linewidth]{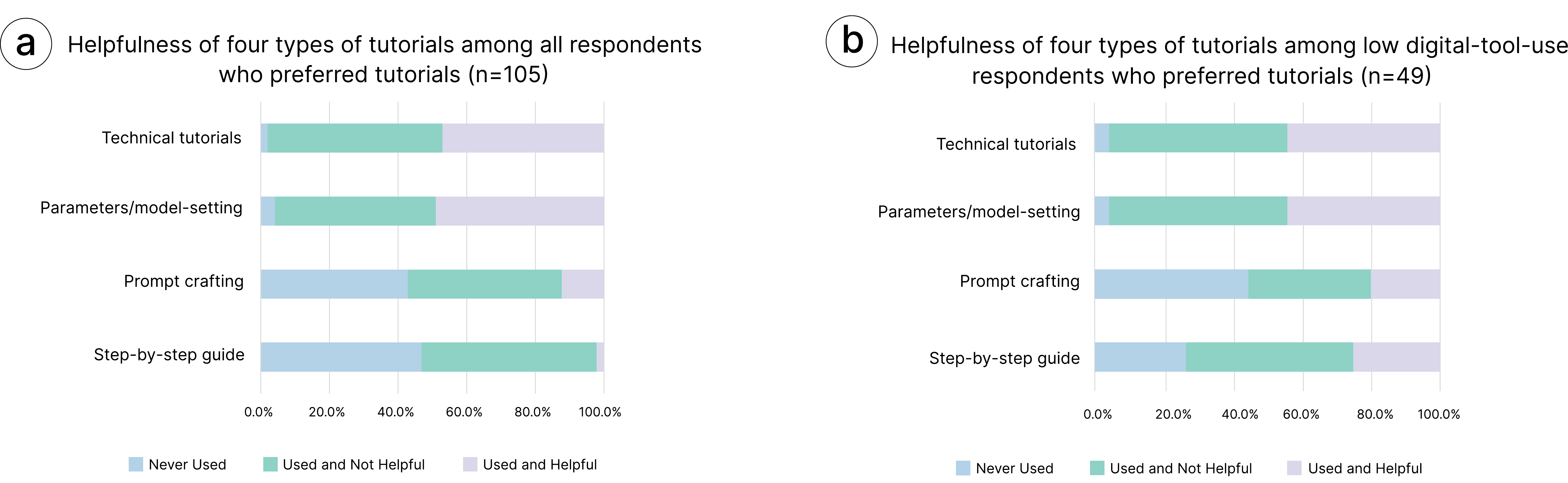}
    \caption{Helpfulness ratings of four GenAI tutorial types among respondents who preferred tutorials, shown all tutorial-preferring respondents (n = 105) (a) and the low digital-tool-use tutorial-preferring subgroup (n = 49) (b). Across both groups, technical tutorials and parameter/model-setting tutorials were the most widely used, with responses split between those who found them helpful and those who did not. The four tutorial types included: (1) step-by-step guides for creating AI images; (2) tips for crafting prompts; (3) parameter/model-setting tutorials (e.g., how to set the CFG scale, sampling steps, sampling methods, and model selection); and (4) technical tutorials explaining behind-the-scenes concepts (e.g., neural networks, diffusion models, and training processes).}
    \label{fig:different}    
\end{figure*}

\subsection{Follow-up Survey}

To examine whether these patterns extended beyond our interview sample, we surveyed how creatives learn GenAI image tools, what terminology-related frustrations they encounter, and how they use existing resources.

\textbf{Survey Design:} The survey comprised twelve questions. It began by collecting background information such as demographics and experience with creative practice and GenAI image tools. To investigate approaches to engaging with GenAI image tools, respondents were asked: (1) What is your most preferred way to learn about GenAI image creation tools (e.g., DALL·E, Midjourney, Stable Diffusion, Artbreeder, Runway ML, or others)? and (2) If you don't prefer using tutorials, what is the main reason? The respondents who preferred tutorials then rated their prior experience with four types of tutorials illustrated with examples: step-by-step guides, prompt-crafting tips, parameter/model-setting instructions, and technical explanations. Finally, respondents were asked if there were any additional tutorial formats they had found useful. Key results of the approaches questions are shown in Figure \ref{fig:surveyfindings}. Respondents were recruited through our local university and community contacts and online platforms such as Reddit and Twitter and were entered into a raffle for a \$50 Amazon gift card.

\textbf{Respondent Demographics and Backgrounds:} We initially received 220 responses, and then manually inspected the response data and removed those that were spam, nonsensical or incomplete. This process left us with a total of 159 responses from participants across 13 countries, including the United States (44.7\%), Canada (34.6\%), the United Kingdom (13.8\%), and other countries (6.9\%). Respondents had diverse art education backgrounds, with the majority (64.1\%) holding a bachelor's degree or higher in art or a related field, 30.2\% having some art education, and 5.7\% having no formal art education.

\textbf{Results:} Building on prior work on creatives' difficulties in understanding AI systems \cite{15,51} and our interview findings, we used the survey to contextualize how creatives currently attempt to make sense of GenAI image tools. We compared a low digital-tool-use subgroup (i.e., respondents who never, rarely, or sometimes use digital tools in their art practice; $n=80$) with the full sample ($n=159$).

Survey responses showed that creatives most commonly relied on third-party tutorials (52.8\%), such as YouTube videos or online courses, followed by self-experimentation (29.6\%), while official tutorials provided by tool platforms (e.g., DALL·E or Midjourney) accounted for only 13.2\%. Despite the prevalence of tutorials, many respondents described challenges when using them. Difficulty understanding terminology was the most frequently reported barrier (37.0\% overall), with low digital-tool-use respondents disproportionately affected. As shown in Figure~\ref{fig:surveyfindings}b, 64.5\% of low digital-tool-use respondents avoided tutorials due to complex terminology, compared to 37.0\% across all respondents. A notable proportion of respondents also reported avoiding tutorials because they preferred self-experimentation over structured guidance (29.6\%).

To examine how tutorials function when they are used, we analyzed respondents' experiences with different types of tutorial content. As shown in Figure~\ref{fig:different}, technical tutorials explaining AI concepts (e.g., neural networks, diffusion models, training processes) were widely used across both groups (98.0\% of all respondents and 95.9\% of low digital-tool-use respondents reported having tried them), whereas basic step-by-step guides and prompt-crafting tips were less commonly used. However, respondents reported mixed experiences with technical tutorials: roughly half in both groups found them unhelpful, suggesting that although creatives were genuinely interested in these resources and made efforts to use them, their effectiveness may be hindered by the complexity of the terminology and underlying concepts.

Some respondents reported dissatisfaction with tutorials that instructed them to replicate settings without providing conceptual explanations of their underlying meaning or purpose. Respondents wanted to understand what each setting represents and how it impacts the output. One respondent commented, \textit{``Copy and paste the settings, like all the settings in Latent Modifier Integrated? What does that even mean?''}. 

In summary, third-party tutorials emerged as the preferred way that creatives (esp. in the low digital-tool-use category) sought to learn GenAI tools, yet many also described them as difficult to follow because of jargon. Technical and parameter-setting tutorials were the most widely used, though opinions were sharply divided between finding them helpful and not helpful. These patterns suggest that creatives often want to reason about how GenAI systems behave, but existing resources do not reliably support that conceptual understanding, especially when terminology dominates.

\section{Exploring Creatives' Experience with Structured Guidance and Self-Experimentation}

Our formative interview and survey showed that creatives commonly learned GenAI image tools through third-party tutorials or self-experimentation, while technical terminology remained a key barrier to conceptual understanding.

Building on these findings, we sought to examine how creatives respond to simplified structured guidance in practice, relative to self-experimentation. Existing tutorials are often highly technical \cite{nvida, jay} or tightly coupled to specific tools \cite{nightcafework}, making it difficult to isolate how simplified conceptual guidance shapes users’ understanding and experience. 

To address this, we developed a research probe to isolate simplified structured guidance without technical jargon and to elicit creatives’ reactions, expectations, and concerns in use \cite{boehner2007}. Using this probe, we conducted a comparative structured observational study \cite{comparative} with 17 visual artists and hobbyists, followed by interviews. This qualitative approach systematically varies conditions to support direct comparison, not to evaluate tool effectiveness, but to elicit reflection on how participants formed mental models through structured guidance or self-experimentation.

\subsection{Exploring Structured Guidance with a Research Probe}

\subsubsection{Initial Design Exploration and Expert Feedback Sessions}

We started developing our research probe by prototyping plain-language, visual explanations of core concepts (e.g., patterns, semantic meaning, attention). To refine the level of detail and pedagogical clarity, we conducted feedback sessions with eight experts: six artists (3M, 3F) specializing in digital illustration, visual arts, and traditional painting, and two computer vision researchers (2M) with 3 and 10+ years of experience, respectively.

\begin{figure*}[tb]
  \centering
    \includegraphics[width=0.9\linewidth]{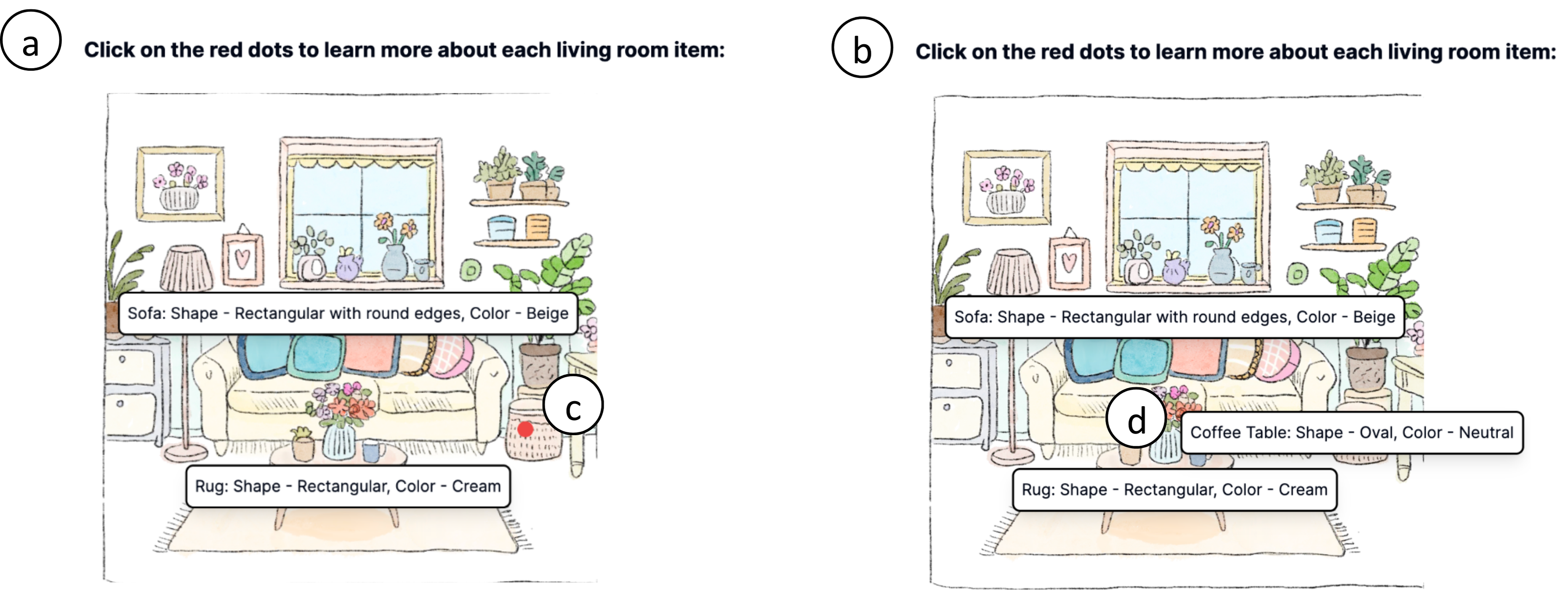}
    \caption{Early design exploration of our research probe, where users could click on red dots to reveal object attributes in a living-room scene. In (a), users view tooltips for the sofa and rug, and in (b), an additional coffee-table tooltip appears after further interaction. While this approach illustrated how AI tools might label objects, expert feedback noted that it risked oversimplifying AI learning. GenAI does not rely on identifying single objects, but on statistical patterns across millions of training images. This realization prompted us to pivot toward designs that emphasize large-scale concept learning in the image generation process.}
    \label{fig:initial}
\end{figure*}

\textbf{Simplifying Concepts Without Oversimplifying AI:} The artists engaged with the prototype by exploring its text-to-image and style transfer features, experimenting with prompt modifications and assessing how style changes influenced outputs. They found the conversational format explanations easy to follow and engaging, and suggested adding options to vary prompt elements and adjustable parameters (e.g., style strength) to better support comparison alongside the explanations. The two computer vision experts focused on ensuring the tutorial did not misrepresent the generative process. For example, they noted that an early prototype (Figure \ref{fig:initial}) that revealed object attributes via clickable dots risked implying object-by-object reasoning, rather than pattern learning across large datasets. Based on this feedback, we revised the tutorial to more explicitly emphasize large-scale concept learning in image generation.

\textbf{Realistic vs Comic-style Images:} The technical experts recommended using realistic images to emphasize the authenticity of the demonstrated processes, noting that comic-style images might give the impression of a hypothetical process. In contrast, the artists focused on the source of images and the conversational style used in the tutorial. They also preferred comic-style images as they believed these better aligned with the tutorial's narrative and kept the content engaging. Comics have been widely used in scientific communication to combine visuals, text, and narrative flow, making abstract concepts more approachable and memorable \cite{wang_data_nodate, cbc_how_2025}. Prior work also shows that comic-style explanations can improve engagement and comprehension compared to text-only or traditional formats \cite{compare2019, farinella}. Based on this feedback and prior research, we then chose to retain the comic-style images.

Based on the feedback from these sessions, we iterated Peek-Box's design, emphasizing clear, easy-to-understand explanations of underlying image-generation mechanics, interactive elements, and a playful, comic style. Drawing on literature as well as from our formative study suggesting that users prefer simpler, broader explanations over detailed accounts of specific models \cite{read, lombrozo}, we shifted our focus to generalized key concepts in the image generation process to help users across platforms grasp foundational basics.

 \begin{figure*}[tb]
    \centering
    \includegraphics[width=0.7\linewidth]{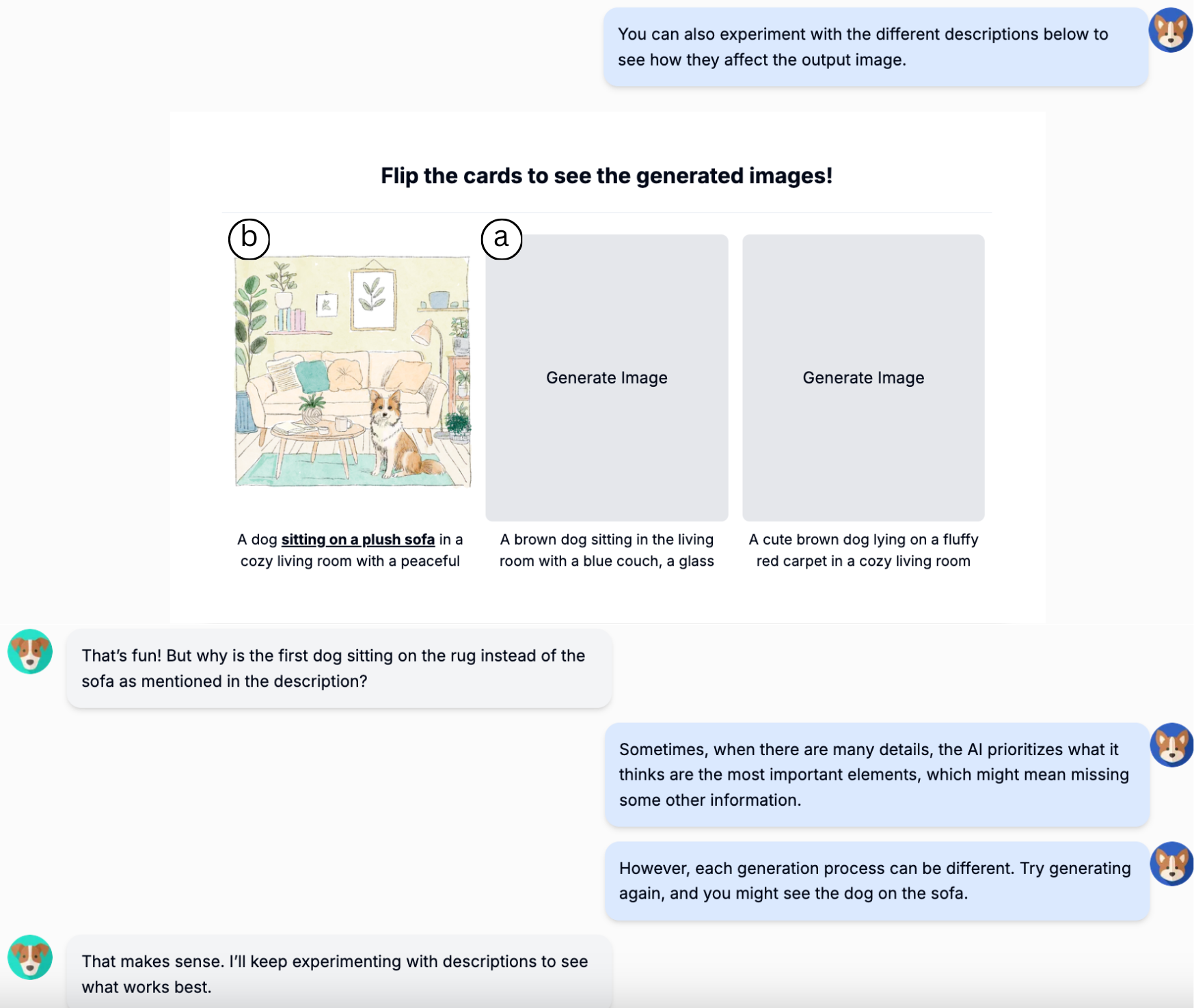}
    \caption{An example of the Text-to-Image tutorial. The interactive design allows users to click through the explanation interactively. Here, we aim to show the effect of different prompts. Each card corresponds to the prompt displayed below it. Before users click on a card, all cards display ``Generate Image'' (a). After clicking, the card flips to reveal the image (b) corresponding to the prompt displayed below the card. Since the first image (b) is missing some information from the prompt, we then introduce the AI's attention mechanism.}
    \label{fig:attension_image}        
\end{figure*}

\subsubsection{Final Probe Structure}

Peek-Box was a structured research probe presented in a conversational format across three contexts: text to image, style transfer, and sketch to image. In the text-to-image context, participants first generated an image from a simple prompt such as ``A dog is sitting in the living room,'' and then progressed through short, step-by-step explanations of how the system might form concepts like ``living room'' by learning patterns from labeled training images. Peek-Box also included lightweight interactive elements, such as flip cards that reveal outputs for specific prompts (Figure~\ref{fig:attension_image}).

Prior work shows that artists often evaluate results first and then infer what model behaviors, prompt components, or parameter choices might have produced them \cite{kim2024, Palani2024}, and that this backward reasoning supports mental model formation in opaque systems \cite{limdey2010, kulesza2013, norman1983}. To align with this reasoning pattern, Peek-Box presented outputs before explanations. Across contexts, the probe used contrastive examples to illustrate how changes in prompts and settings can shift outputs and to surface cases where results diverged from intended details, supporting more realistic expectations. The style transfer and sketch to image contexts extended this framing by allowing participants to adjust a style intensity or creativity setting and observe trade-offs between constraint and variation.

\subsection{Participants} 

We recruited 17 participants (6F/10M/1GNC, gender non-conforming) from various professions, including artists, graphic designers, web designers, students, civil engineers, executive directors and photographers. Most participants (14 out of 17) had some experience with GenAI image tools (Table \ref{tab:over_participants}), but none of them were ML experts. Participants included both self-identified artists and art hobbyists; across the sample, Gen AI was used primarily for exploration and early-stage ideation. Recruitment was conducted through advertising within our university, personal connections, social media advertising, and snowball sampling. 

\begin{table*}[t]
    \centering
    \caption{Participants come from varying backgrounds, with 11 out of 17 being visual artists, and the rest of them (P4, P5, P6, P8, P12, and P16) are art enthusiasts.}
    \small
    \begin{tabular}{l l l l} 
        \toprule
        \textbf{Participant} & \textbf{Gender} & \textbf{Background} & \textbf{Prior experience using Gen AI image tools} \\ 
        \midrule
        1 & M & Graphic Design & 3-5 times \\ 
        2 & F & Interaction Design & 3-5 times \\
        3 & M & Photography & 6-9 times \\
        4 & M & Geography & 10 times or more \\
        5 & M & Engineering & Never \\
        6 & F & Computer Science & 1-2 times \\
        7 & M & Performing Arts & 6-9 times \\
        8 & F & History & Never \\
        9 & F & Information Visualization & 3-5 times \\
        10 & F & Architecture & 1-2 times \\
        11 & F & Interactive Art & 10 times or more \\
        12 & M & Education & Never \\
        13 & M & Business Management & 10 times or more \\
        14 & M & Visual Art & 3-5 times \\
        15 & M & Graphic Design & 10 times or more \\
        16 & M & Computer Science & 10 times or more \\
        17 & GNC & Fine Art & 3-5 times\\
        \bottomrule
    \end{tabular}
    \label{tab:over_participants}
\end{table*}


\begin{figure*}[t]
    \centering
    \includegraphics[width=0.9\linewidth]{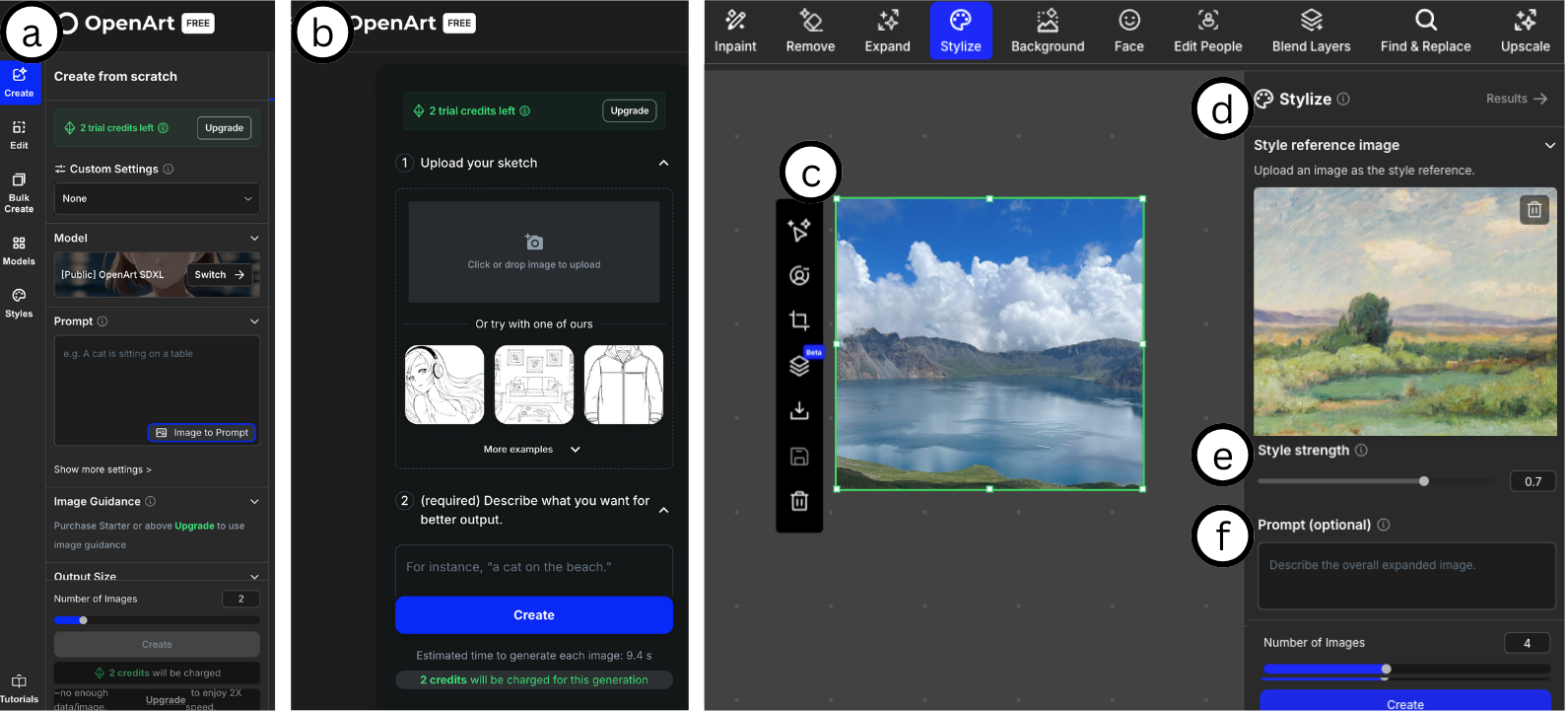}
    \caption{Openart AI interface: (a) Interface where participants interacted with the text-to-image tasks; (b) Interface for sketch-to-image task. (c) Interface for the style transfer task, where participants upload the content image first, after which the stylize section (d) appears, allowing them to upload the style image. Additional parameters, such as style strength (e) and an optional prompt (f), can be adjusted to further customize the output.}
    \label{fig: interface}
\end{figure*}

\subsection{Study Design and Procedure} 

\textbf{Choice of Application} We selected \textit{OpenArt AI} \cite{noauthor_create_nodate} for its clear, modular interfaces for text-to-image, style transfer, and sketch-to-image tasks (Fig \ref{fig: interface}) and its minimal reliance on complex parameter tuning, reducing potential confusion. We chose a different platform intentionally to avoid tying findings to a specific model or UI and to focus on conceptual understanding that can transfer across GenAI image tools.

\textbf{Study Design} Insights from our formative study revealed that creatives most commonly adopt one of two strategies to learn GenAI tools: using third-party tutorials or relying on self-experimentation through trial-and-error. These strategies are not mutually exclusive. As \textit{Situated Learning Theory} \cite{lave1991situated} suggests, learning unfolds through ongoing practice, and creatives may alternate between guidance and experimentation. We therefore examine these approaches as task-adjacent supports rather than separate stages. To reduce sequencing effects, we varied the order of tasks and approaches across participants using a \textit{Latin Square} schedule, yielding 16 possible task and approach sequences. Across the session, participants experienced both approaches, allowing them to compare and reflect on their experiences.

\begin{figure*}[t]
    \centering
    \includegraphics[width=0.9\linewidth]{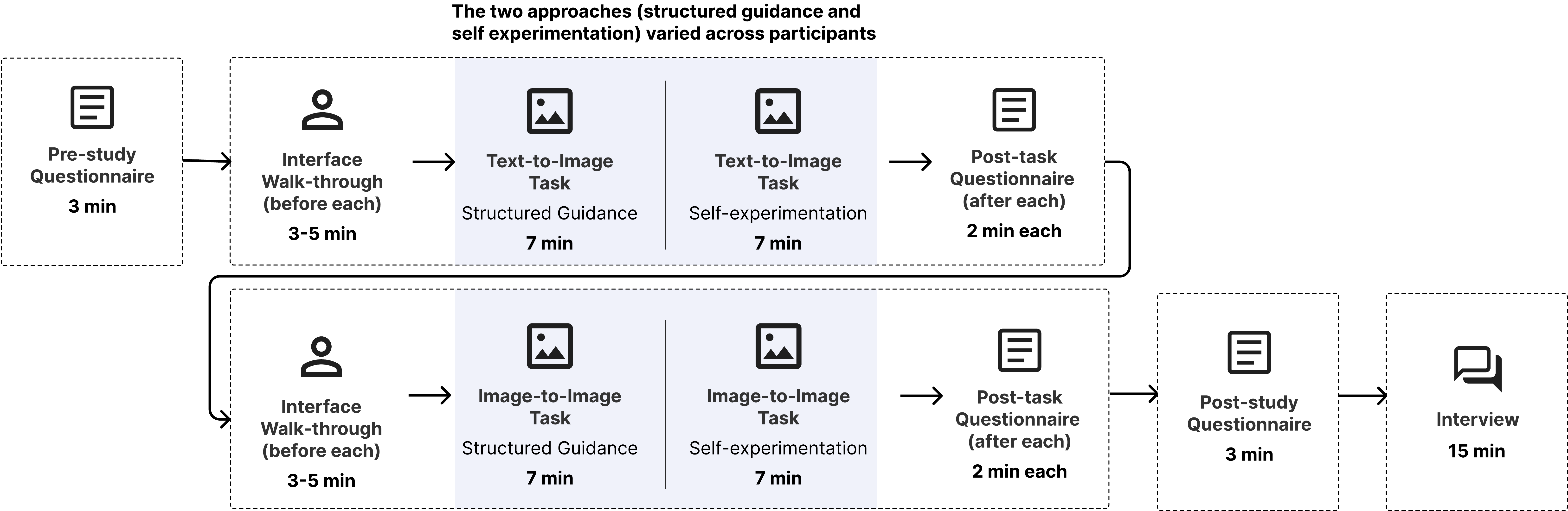}
    \caption{User study session procedure. Participants worked through two text-to-image tasks and two image-to-image tasks, completing each task once with structured guidance using the research probe and once through self-experimentation. By experiencing both approaches, participants could compare them directly and reflect on their experiences, following the comparative structured observational study method \cite{comparative}.}
    \label{fig: userstudyprocess}
\end{figure*}

\textbf{Study Procedure} Sessions began with a brief introduction, followed by informed consent. Participants then completed a pre-study questionnaire capturing demographic information, prior experience with GenAI image tools, and general perceptions of these technologies (see Fig \ref{fig: userstudyprocess} for the user study session procedure).

Participants then completed four tasks using a think-aloud protocol. The tasks aligned with the tutorial content and covered text-to-image generation, sketch-to-image conversion, and style transfer. Before each task, a researcher provided a short interface walkthrough of about five minutes to ensure participants could locate the features required for the task. For the two text-to-image tasks, each participant completed one task with structured guidance and one with self-experimentation. They followed the same pattern for the two image-to-image tasks. A researcher observed participants' interactions with each approach throughout the session. Each task lasted up to seven minutes and was followed by a short post-task questionnaire capturing immediate impressions such as satisfaction and confusion.

After completing all four tasks, participants completed a post-study questionnaire and a semi-structured interview that asked participants to reflect and compare their task experience with different approaches and their perceptions of the structured guidance.

Each session lasted about an hour, and participants received a \$20 Amazon gift card. The study received approval from our institution's research ethics board.

\begin{figure*}[t]
    \centering
    \includegraphics[width=0.8\linewidth]{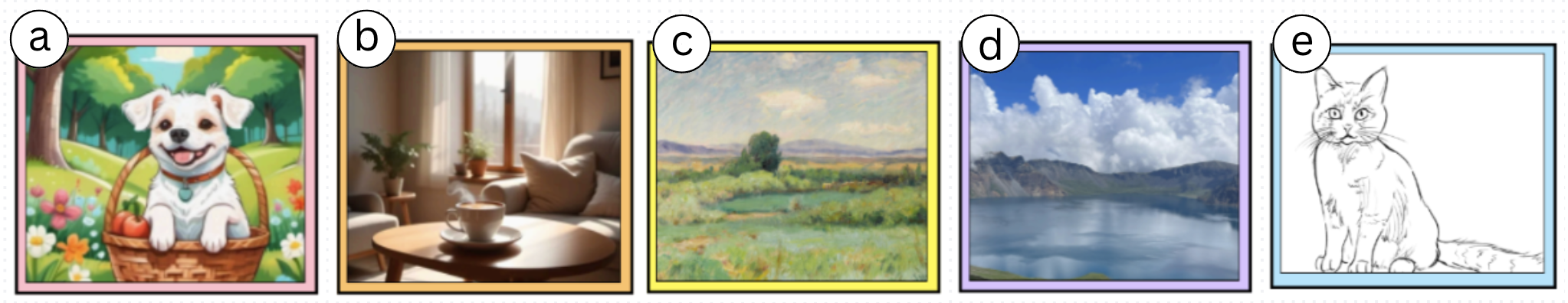}
    \caption{Images used in the study tasks. (a) and (b) for text-to-image, (c) and (d) for style transfer, and (e) for sketch-to-image generation.}
    \label{fig: tasks}
\end{figure*}

\subsection{Data Collection and Analysis}

We gathered screen and audio recordings of the think-aloud study, images created by each participant and collected responses from questionnaires. We also collected the researcher's handwritten notes taken while observing the participants.

\textbf{Interpretive Analysis}: The audio recording of the study was transcribed. We used an \textit{inductive analysis} approach \cite{strauss1994grounded} to explore emerging themes from the on-screen interactions, think-aloud, notes, and follow-up interviews. Next, we used axial coding to organize participant responses around three key dimensions: (1) how participants conceptualized the image generation process, (2) how participants perceived and used the structured guidance in the tutorial, and (3) how they interpreted unexpected or undesired outputs produced. 


\textbf{Descriptive Analysis of Questionnaire Responses:} We analyzed post-task questionnaires to capture participants' self-reported confusion, confidence, enjoyment, and satisfaction for each task on 5-point Likert scales (higher scores indicate greater intensity or agreement). Tutorial helpfulness was rated from Strongly Disagree (1) to Strongly Agree (5). Following the comparative structured observation method, we treat these ratings as a secondary context for interpreting participants' qualitative reflections, not as measures of task performance.

\section{Results}

Our key results focus on the initial mental models creatives formed about GenAI image tools, differences in the explanations and next-step strategies participants articulated with structured guidance versus self-experimentation, and the tensions participants navigated between conceptual clarity and creative autonomy. 

\subsection{Misconceptions and Initial Gulfs in Understanding GenAI Image Generation}

One pattern in our study was the disconnection between participants' reported confidence and their actual understanding of how GenAI works. While a majority (10/17) expressed confidence in the pre-study questionnaire—agreeing with the statement, ``\textit{I understand the inner working mechanism of GenAI image tools}''---their subsequent explanations revealed widespread misconceptions that reflected a classic gulf of evaluation \cite{norman1983}, the difficulty in interpreting what the system is doing or why. Even when participants were primarily focused on getting usable outputs rather than understanding AI in depth, these beliefs still shaped how they explained unexpected results and what to try next. Many believed that AI simply \textit{``searches for''} and \textit{``combines online images''} based on the prompt (P1, P2, P4, P14, P15) or that it retrieves results from a pre-existing dataset (P3, P7, P10, P13, P16). The first view treats GenAI as an advanced search engine, while the second conflates training data with generated outputs. This is consistent with \citet{norman1983}'s view that mental models do not have firm boundaries; when mechanisms are opaque, people often reuse familiar system analogies.

These evaluation difficulties also contributed to a gulf of execution \cite{norman1983}, a mismatch between what users intend to do and what actions they believe the system affords. In several cases, when participants could not interpret model behavior, they defaulted to prompt-centric action strategies. For example, when the output fell short of expectations, participants blamed the prompt (and indirectly themselves) rather than recognizing model limitations or the inherent variability in GenAI processes. P1 explained an unexpected output by saying, \textit{``\ldots{}probably that could be the prompt that I chose, maybe it kind of diverts from what I originally wanted. So, I don't think it's AI's problem.''(P1)} P4 encountered a similar problem and commented:\textit{``It's all about the prompt\ldots{} not AI, just prompt, AI has tried its best to answer my instructions.''(P4)} Finally, P13 also emphasized the role of the prompt when describing what he thinks about how AI works: \textit{``The refinement comes in if the response is too simple or incorrect. Quite often, it's just bad language in the way of the prompt.''(P13)}

Together, these accounts show how initial misconceptions influence both how participants evaluated unexpected outputs and what actions they believed were available to improve results.

\subsection{Perceptions of Structured Guidance}

Most participants reported that the simplified structured guidance helped them better understand how AI interprets inputs and generates output, for both text-to-image (16/17) and image-to-image tasks (15/17). More importantly, some participants' reflections shifted from prompt-only explanations toward reasoning about how the system prioritizes and interprets prompt elements. Several participants reported becoming more specific in their prompts (P1, P5) and better understanding \textit{why} outputs sometimes differed from their expectations (P6, P10). For example, when asked about any surprising outputs in the process, P6 said, \textit{``I think all of them [the output images] kind of not surprised that much. Yeah, I think AI chooses what's most important in the sentence and highlights those specific elements.''(P6)} Similarly, P10 observed that the model missed her \textit{``zoom in''} instruction, noting that \textit{``maybe, as the tutorial said, it prioritized some specific words while missing other information.''(P10)} Before interacting with the tutorial, P10 believed that \textit{``the more detail that you input [in the prompt], [the] more detail the photos come out.''} Afterwards, she realized what was actually going to be prioritized:\textit{``It was new information for me based on the tutorial\ldots I will be aware of what I [say] to AI.''(P10)}

\textbf{Post-Task Reflections on Self-Experimentation vs. Structured Guidance:} Across tasks, participants reported slightly lower confusion ratings (structured guidance: M = 2.33, SD = 0.92 vs self-experimentation M = 2.68, SD = 1.22) and slightly higher satisfaction (structured guidance: M = 3.91, SD = 0.78 vs self-experimentation M = 3.36, SD = 0.99) when structured guidance was available compared to self-experimentation. Confidence ratings were comparable with self-experimentation, while enjoyment showed little difference. We use these ratings to contextualize participants' qualitative reflections rather than to evaluate the effectiveness of the research probe. The patterns we observed suggest that participants experienced the tutorial as reducing uncertainty and improving their sense of output quality without greatly altering confidence or enjoyment. For example, P7 who started with self-experimentation and later tried the tutorial, felt that the tutorial, \textit{``would [have] been more supportive to enhance learning\ldots{} at first''} and noting that during the self-experimentation phase \textit{``it was more confusing to me \ldots{} trying to make sure the output matches what I had on the PDF [task description provided in the session] was kind of frustrating.''(P7)} 

Participants also used the post-task reflections to compare not only how they felt but also how they decided what to try next. For example, P3 began his first task through self-experimentation. When asked how he thought the AI interprets his prompt to generate the dog image (Task A), he explained, \textit{``AI will go around the web to look for pictures of dog.''} After completing all the tasks and viewing the tutorial, which highlighted that with long prompts the model may overlook some details, he noted that he wanted guidance on how detailed his prompt should be. He reflected that \textit{``including too much information could overload the AI and could turn a person or an individual into a mess.''(P3)}

\subsection{Tensions in Navigating Guidance vs. Maintaining Creative Autonomy in Image Generation}

Across the study, participants' experiences revealed several tensions in balancing structured guidance with self-experimentation during AI-assisted image generation. While many participants described the structured guidance as helpful and reported being more satisfied with some images produced with it, their reflections and behaviors suggested a more nuanced relationship between guidance, confidence, and creative freedom. We first unpack the tension around creative autonomy, then discuss two related observations around confidence/confusion and ethical reflection.

\textbf{Creative Autonomy vs. Structured Support} Although participants often reported higher satisfaction with their outputs, and most also reported that the guidance helped them make sense of how the AI interprets prompts and generates images, four participants (P1, P2, P5, P13) resisted fully engaging with the tutorial, fearing it might limit their creative instincts. We interpret this not as a general preference, but as a design tension: guidance can support understanding while still feeling prescriptive for some creatives. As P2 explained, \textit{``I just kind of go off my own intuition rather than going off a tutorial. Most of the time, I don't really like to be guided when I'm doing things that are, like, more creative.''(P2)} This sentiment was echoed by P5, who rated one task as only slightly enjoyable because he preferred to work without external guidance. P13 expressed a similar preference, describing self-experimentation as \textit{``following my own guidance from inside''}, yet he also asked when the content might be available to others, noting that he would recommend it to newcomers in his field as a way to learn basic AI concepts.

At the same time, other participants welcomed structure or questioned the relevance of understanding how the AI works. P7 viewed guidance, whether from external tutorials or internal intuition, as a valuable resource that supported their creative process. In contrast, P11 was skeptical about the usefulness of background knowledge, stating, \textit{``So I [am] never\ldots interested in knowing the background [technical details] behind this AI tool, I just use it. It's because, like, knowing the background doesn't mean you can get better output.''(P11)}

\textbf{Confidence and Confusion as Context} Although participants with structured guidance reported lower levels of confusion in their post-task ratings, several (P1, P2, P3, P5, P13) appeared reluctant to acknowledge confusion during the study, even when they were visibly struggling. For instance, P1 rated himself as ``not confused'' in the post-task questionnaire for the Sketch-to-Image task, even though the first few outputs did not even include a cat, which was central to the task. In the follow-up interview, when asked about missing the cat and still rating himself as ``not confused'', P1 then admitted the task had been ``a little bit confusing''. Similarly, P2 expressed uncertainty multiple times during the text-to-image tasks, saying, \textit{``I don't know why\ldots''(P2)} when the AI-generated output did not meet her expectations, but rated herself as ``not confused'' in the post-task questionnaire.

Prior work shows that people often apply social norms from human interaction to computers \cite{37}. In collocated settings, displaying low confidence can be read not just as uncertainty but as reduced competence or motivation \cite{40}. Thus, confidence can function as a socially shaped performance of competence in HCI \cite{37,40}. In some cases, participants desired to be perceived as confident during the study. For instance, P3's task attempt and generated image for Text-to-Image Task B was not close to the provided image but he still rated himself as being ``confident'' in completing the task and stated, \textit{``I was confident before I got into trouble, but I was confident.''(P3)} P8 similarly demonstrated strong confidence during the Text-to-Image Task A, even when the results did not align with her prompts. She rated herself as ``confident''(4) and believed she could \textit{``eventually get the image right''} if she kept trying.

\textbf{Emergent Ethical Reflections} Structured encounters with AI tools sometimes triggered reflections on training data and copyright. Some participants (P4, P10, P14, P17) did not have any concerns at all due to their non-commercial use of the AI-generated output. Interestingly, P17 assumed (wrongly) that AI models are trained on open-source images and that's why he was not worried. 

While some participants dismissed concerns, others (P2, P6) stressed the importance of attribution, consent, and transparency. P2 and P6 did raise some ethical concerns about AI's use of copyrighted material: \textit{``I feel like it's really important to me as an artist because if I put my work out there, I don't really want it to be used for AI and not be attributed to it, especially if I'm putting a lot of effort into my work and producing it. Just having it stolen by AI and nobody even knows \ldots that's not what I want as an artist.''(P2)} P2 expressed the wish for AI tools to get explicit permission from artists before using their work, adding, \textit{``It makes it so much better if they actually get permission from us because, like, everything has copyright nowadays, and our art should have copyright too.''(P2)} P6 spoke to the importance of knowing more about the source of the training data as that would increase trust in AI, noting: \textit{``I think for me it [knowing the source] does help. But I know it still does not help some people because people are just scared of AI in general.''(P6)}

\section{Discussion}

Building on the qualitative patterns reported above, we interpret our findings as showing how participants in our study formed mental models when engaging with GenAI image tools, and how simplified structured guidance could both support and constrain AI literacy initiatives for creatives. While prior work has focused on prompt engineering \cite{promptify} or interface affordances and visualization \cite{chung_promptpaint_2023,prompt_this}, we foreground creatives' learning pathways, highlighting the tension between structured guidance and creative autonomy. Table~\ref{tab:designtensions} summarizes three design implications: aligning explanations with creative goals, providing in-context support that preserves flow, and offering adaptive guidance for diverse literacy needs. Together, these insights point toward user-centered tutorials that support both conceptual understanding and creative freedom.

\begin{table*}[t]
\centering
\caption{Summary of interpretive implications drawn from the tensions observed across our studies. Please note that these are not prescriptive design guidelines, but reflections on how learning support intersects with creative practice.}
\small
\setlength{\tabcolsep}{6pt}
\renewcommand{\arraystretch}{1.8}
\begin{tabular}{p{5.3cm} p{4.2cm} p{6.0cm}}
\toprule
\textbf{Observation} & \textbf{Tension}  & \textbf{Implication for supporting AI literacy} \\
\midrule

Creatives found structured guidance useful, especially for reasoning about unexpected results 
& Recognizing limitations vs maintaining creative momentum
& Surface plausible limitation cues for unexpected results; use unexpected outputs as learning examples \\

Creatives often valued coherent, actionable guidance over technical fidelity
& Who defines a ``good'' guidance and who the guidance serves
& Prioritize the target users' creative goals; keep technical depth optional \\

Creatives' need for guidance varied by depth and by moment in the creative process
& One-size guidance vs.\ creating in active use 
& Offer on-demand, in-context guidance that lets users choose depth and timing of guidance  \\

\bottomrule
\end{tabular}
\label{tab:designtensions}
\end{table*}

\subsection{Interpreting Findings through the Lens of Mental Models}

Our findings can be interpreted through the lens of mental models \cite{norman1983}.  With simplified structured guidance, participants more readily articulated causal accounts of system behavior, such as how the system prioritizes certain prompt terms or decides what output to generate from a given input. This suggests that structured guidance can help narrow the gulfs of execution and evaluation by supporting the formation of more explicit and reasoned mental models. In contrast, self-experimentation immersed participants in concrete experience and reflective observation, giving them a lived sense of how the system responds as they iterate. This approach supports more intuitive, experience-based mental models that emerge through trial and error. While these models can offer effective practice-tested strategies, they are often fragmented or incomplete, leaving participants prone to surprise, misattribution, or overgeneralization. We observed that structured guidance can reduce uncertainty early in the learning process and help users reason more confidently about the system, yet the same structure can also shape what participants attempt next, sometimes constraining improvisational momentum.


\subsection{Reflecting on the Paradox of Using Structured Guidance}

Although most participants found structured guidance helpful in understanding AI, some still preferred self-experimentation, citing concerns that pre-task guidance would diminish their creative expression.

This behavior echoes the \textit{active user paradox} \cite{johnuserparadox}, where users bypass tutorials in favor of direct engagement, often overestimating their ability to learn through self-experimentation \cite{trial}. In our context, this tendency also reflects  \textit{production bias} \cite{johnuserparadox}: participants preferred to maintain creative momentum rather than step away from making. Prior work similarly shows that creatives assess new tools by how well they fit their existing work patterns and preserve a sense of smooth creative flow, not just by how powerful they are \cite{1960s}. However, unlike traditional cases of the active user paradox observed in software learning\cite{kimia, marc, field}, reluctance to follow tutorials here was not just about self-reliance but a desire to maintain creative autonomy with AI tools. This positions creative autonomy as a domain-specific expression of production bias, where the drive to ``keep creating'' can conflict with opportunities for deeper understanding, revealing a complex relationship between learning and freedom in creative work.

 This tension between guidance and creative flow resonates with previous research \cite{amir2024, 16, littension, littension2} and presents new questions for the HCI community to consider: How can we support the effective use of AI-based creative tools without making creatives feel constrained? In our study, we also observed that creatives tend to approach AI understanding selectively, preferring to learn about how AI works only insofar as it supports their immediate creative goals, rather than seeking comprehensive technical understanding of the system. This aligns with prior work that advocates for teaching only the aspects of AI that matter to users' tasks \cite{kelley_advancing_2023} and prior work indicating that users with greater AI knowledge are more likely to adopt partnership-oriented mental models, whereas less experienced users often treat AI as a tool to be directed \cite{rezwana2025mentalmodels}.

\subsection{Addressing Misconceptions and Misunderstandings about GenAI}

A key finding with implications for design was the gap between participants' understanding of how GenAI works and its actual capabilities, which could lead to frustrations and self-blame when the AI failed to meet their expectations \cite{anjaliiui} and propose unrealistic or unscalable ideas \cite{aiwildcard}. Previous research suggests that educating users about AI's capabilities \cite{guidelines} or using onboarding tutorials to demonstrate AI's limitations \cite{novice} could bridge this gap. Our participants echoed this, with P10 and P7 noting that flawed results could be useful for learning---an idea supported by prior education research \cite{productive}. When AI tools produce unexpected results, interfaces could offer optional guidance to help users interpret outcomes and suggest the next step, turning unexpected outputs into learning.

These observations raise broader questions about how to design onboarding experiences and AI-literacy supports that align with creatives' mental models. In our expert feedback sessions, we encountered tensions over which aspects of AI literacy to emphasize in the tutorial: technical experts prioritized accuracy and fidelity to underlying models, while artists emphasized clarity, engagement, and narrative coherence. These perspectives reflect two distinct but legitimate approaches to AI literacy but neither is sufficient on its own. Guidance that are technically correct might be hard to understand and can alienate creative users, while overly simplified accounts risk reinforcing fragile or misleading intuitions. More broadly, effective AI literacy support is not just about presenting correct information, but also about who it is meant to serve and what value is prioritized. Prior work in XAI for arts also emphasizes that an explanation's usefulness cannot be judged purely on technical grounds, since what counts as a meaningful explanation is shaped by an artist's practice and sense of artistic identity \cite{ford2025xaixarts3}. 

\subsection{Empowering Creatives Through a Broader View of AI Literacy}

While the technical-focused strategies could enhance usability, they do not equip users with the conceptual understanding needed to critically engage with AI systems. Prior work has shown that many users of GenAI tools lack foundational AI literacy \cite{static, Burgsteiner, ghallab}, and our study reinforces this finding. We observed that misconceptions about how AI works (or does not work) not only shaped the creative strategies of participants but also influenced their views on how they evaluated AI from an ethical perspective, which is another core competency of AI literacy. 

This limited understanding has broader implications beyond individual user experience. When creatives are not AI-literate, they often approach these tools like any other technology, relying on self-experimentation. While this approach can provide immediate feedback, giving users a feeling of progress without context switching needed \cite{trial}, such self-experimentation could inadvertently increase the computational footprint of the industry. This connects directly to another ongoing concern about the environmental and energy costs of GenAI \cite{roy, jesse}. HCI research on tutorial designs can help mitigate this by minimizing unnecessary experiments and reducing resource usage. 

Finally, future work can explore ways to surface creatives' responses to core competencies of AI literacy that extend beyond technical comprehension. While ethics was not a primary focus of our study, participants' comments about copyright attribution suggest that creatives seek not only functional understanding but also critical awareness. Prior interview work has examined these ethical concerns in depth, such as creatives' attitudes toward their work being used as training data \cite{gero2025}. From this perspective, our probe can be seen as an exploration toward ``critical AI literacy'': flawed mental models often lead to misinformed user behavior \cite{kulesza2013}. Building on this, future learning resource designs could more directly support ethical reflection by clarifying how training data relates to outputs and by surfacing questions around provenance, authorship, and responsible use. Future studies could also examine how creatives' AI literacy develops over time, including how they plan their interactions with AI, monitor whether the tool supports their goals, and reflect on how AI shapes their creative decisions during collaboration with GenAI \cite{sidra2026generative}.

\subsection{Limitations}

Our study has several limitations. First, our formative survey relied on 159 self-reported responses from artists and hobbyists who participate in online communities, introducing potential biases like self-selection and social desirability. Second, our lower digital-tool-use subgroup was defined based on self-reported frequency of using digital tools in art practice, which only imperfectly captures technical background. Future work could use more precise measures, such as formal education, programming experience, or self-reported AI familiarity, to better capture variation in technical expertise among creatives. Additionally, while our analysis highlighted issues with tutorial complexity, it is also possible that some users struggled not because of the content itself but because effective tutorials can be difficult to find. Since our study suggests that creatives may prioritize achieving their creative goals, it may be valuable to explore project based AI literacy support. Future work could explore personalized tutorials that adapt to users' goals and prior knowledge, aligning with prior work suggesting that AI literacy needs can be highly individualized \cite{ford2025xaixarts3}. While we touched on ethical concerns raised by creatives, our study primarily focuses on the conceptual understanding side of AI literacy. Future work could examine how creators' ethical concerns intersect with and are shaped by their conceptual understanding of AI systems.

\section{Conclusions}

This paper uses the lens of mental models to understand how creatives build AI literacy when working with GenAI image tools. We find that simplified structured guidance can support creatives' reasoning about GenAI, yet many still prefer self-experimentation to protect their creative autonomy. These findings surface a central tension for HCI: learning pathways that foster conceptual understanding often diverge from those that sustain creative autonomy, so AI literacy support must balance both aims. Effective learning resources should deepen creatives' understanding of GenAI systems while preserving their sense of ownership and autonomy. Beyond usability, our work positions tutorial design as a lever for addressing misconceptions, supporting ethical awareness, and reducing unnecessary trial and error.

\begin{acks}
We thank the Natural Sciences and Engineering Research Council of Canada (NSERC) for funding this research.

\end{acks}

\bibliographystyle{ACM-Reference-Format}
\bibliography{references2}

\end{document}